\documentclass[12pt]{iopart}
\usepackage{iopams}
\usepackage{dsfont} 
\usepackage[dvips]{graphicx}

\newcommand{\half}{\mbox{$\textstyle \frac{1}{2}$}}

\newcommand{\quat}{\mbox{$\textstyle \frac{1}{4}$}}
\newcommand{\tquat}{\mbox{$\textstyle \frac{3}{4}$}}

\newcommand{\cP}{{\cal P}}
\newcommand{\cT}{{\cal T}}

\begin{document}

\title[Conjecture on the analyticity of ${\cal PT}$-symmetric potentials]
{Conjecture on the analyticity of ${\cal PT}$-symmetric potentials and the
reality of their spectra}

\author[C~M~Bender, D~W~Hook, and L~Mead]{Carl~M~Bender$^{1}$,
Daniel~W~Hook$^{2}$, and Lawrence~R~Mead$^{3}$}

\address{${}^1$Department of Physics, Washington University, St Louis, MO 63130,
USA \\{\footnotesize{\tt email: cmb@wustl.edu}}}

\address{${}^2$Theoretical Physics, Blackett Laboratory, Imperial College,
London SW7 2AZ, UK \\{\footnotesize{\tt email: d.hook@imperial.ac.uk}}}

\address{${}^3$Department of Physics and Astronomy, University of Southern
Mississippi, Hattiesburg, MS 39401-5046, USA \\{\footnotesize{\tt email:
lawrence.mead@usm.edu}}}

\begin{abstract}
The spectrum of the Hermitian Hamiltonian $H=p^2+V(x)$ is real and discrete if
the potential $V(x)\to\infty$ as $x\to\pm\infty$. However, if $V(x)$ is complex
and $\cal{PT}$-symmetric, it is conjectured that, except in rare special cases,
$V(x)$ must be analytic in order to have a real spectrum. This conjecture is
demonstrated by using the potential $V(x)=(ix)^a|x|^b$, where $a,b$ are real.
\end{abstract}

\section{Introduction}
\label{s1}
The field of $\cP\cT$ quantum mechanics \cite{R0} has attracted significant
interest in recent years and a large community of active researchers has
developed. This area of study began with the observation that the complex
$\cP\cT$-symmetric non-Hermitian Hamiltonian
\begin{equation}
H=p^2+x^2(ix)^\epsilon\quad(\epsilon\geq0)
\label{e1}
\end{equation}
has a positive real discrete eigenspectrum \cite{R1,R2}. The property of $\cP
\cT$ symmetry is not sufficient to guarantee that the eigenvalues of a
non-Hermitian Hamiltonian are real; $\cP\cT$ symmetry merely ensures that the
{\it secular determinant} ${\rm det}(H-\mathds{1}E)$ is a real function of $E$
\cite{R3}. The eigenvalues of $H$ are the roots of
\begin{equation}
{\rm det}(H-\mathds{1}E)=0,
\label{e2}
\end{equation}
and thus the condition of $\cP\cT$ symmetry implies that the eigenvalues are
either real or come in complex-conjugate pairs. If the eigenvalues of a $\cP
\cT$-symmetric Hamiltonian are all real, we say that the Hamiltonian has an {\it
unbroken} $\cP\cT$ symmetry, but if there are any complex eigenvalues, we say
that the $\cP\cT$ symmetry of $H$ is {\it broken}.

Lacking further information, one would expect (\ref{e2}) to have some complex
roots. Thus, it was surprising to find that the class of $\cP\cT$-symmetric
Hamiltonians (\ref{e1}) has an entirely real spectrum. In Refs.~\cite{R1,R2,R2a}
numerical evidence and detailed perturbative asymptotic analysis was presented to
show that the eigenvalues of the Hamiltonian (\ref{e1}) are real when $\epsilon
\geq0$. A rigorous proof that these eigenvalues are all real was given by Dorey,
Dunning, and Tateo
\cite{R4,R5}.

Using the WKB quantization condition
\begin{equation}
\int_{x_1}^{x_2}dx\sqrt{E-V(x)}=\left(n+\half\right)\pi,
\label{e3}
\end{equation}
where $x_1$ and $x_2$ are the {\it turning points} [roots of $V(x)-E=0$], one
can understand heuristically why the eigenvalues of $H$ in (\ref{e1}) cease to
be real when $\epsilon<0$. The quantization condition (\ref{e3}) requires that
there be a continuous integration contour from $x_1$ to $x_2$. Such a contour
exists for $\epsilon\geq0$, but as soon as $\epsilon$ goes below $0$, the contour
joining $x_1$ and $x_2$ is broken by the existence of a branch cut in the
complex-$x$ plane and there is no longer a finite-length path connecting the
turning points.

The discovery that the eigenvalues of $H$ in (\ref{e1}) are real led to a search
for and subsequent study of other non-Hermitian $\cP\cT$-symmetric Hamiltonians
whose spectra are also real \cite{R6,R7,R8,R9,R10,R11,R12}. We emphasize that
the reality of the eigenspectrum is an unusual property of a complex Hamiltonian
and that many $\cP\cT$-symmetric Hamiltonians do not have entirely real spectra.
For example, while the $ix^2y$ potential studied in Refs.~\cite{R13,R14} has a
real ground-state energy, it has recently been found that some of the other
eigenvalues are complex \cite{R15}.

In this paper we conjecture that analyticity of the potential is a necessary
(but not sufficient) criterion for a non-Dirac-Hermitian Hamiltonian to have real
eigenvalues. This conjecture is based on extensive numerical studies in which we
have found that, except in rare cases, a $\cP\cT$-symmetric Hamiltonian $H=p^2+V(x
)$ does not have a real spectrum if its potential $V(x)$ is not an analytic
function of $x$. An example of such a nonanalytic $\cP\cT$-symmetric potential,
which is discussed in Sec.~\ref{s2}, is
\begin{equation}
V(x)=ix|x|.
\label{e4}
\end{equation}
We show in Sec.~\ref{s2} that this potential has only one real eigenvalue.

A heuristic explanation of the role played by analyticity can be based on the
WKB contour integral in (\ref{e3}). For complex $\cP\cT$-symmetric potentials
the derivation and application of this integral makes explicit use of the
analyticity of the potential. Of course, the potential of a Hermitian
Hamiltonian need not be analytic because its turning points lie on the real
axis. Here, the integral for the WKB quantization condition is unambiguously
taken along the real axis and does not need to be deformed into the complex
plane. By contrast, the turning points for a complex potential are likely to be
complex, and thus the contour for the WKB integral necessarily lies off the real
axis. Giving up Hermiticity forces the quantization condition (\ref{e3}) into
the complex plane and thus introduces the requirement of path independence and
hence analyticity.

The square-well potential studied by Znojil in Ref.~\cite{R6} is a rare example
of a complex nonanalytic $\cP\cT$-symmetric potential having a real spectrum.
This potential evades the conjecture above simply because there are no turning
points at all; for the square-well potential there is no solution to the
equation $V(x)=E$.

This paper is organized very simply: In Sec.~\ref{s2} we examine the exactly
solvable nonanalytic potential in (\ref{e4}) and in Sec.~\ref{s3} we present
numerical results for the class of nonanalytic $\cP\cT$-symmetric potentials
\begin{equation}
V(x)=(ix)^a|x|^b\quad(a,\,b~{\rm real}).
\label{e5}
\end{equation}
In Sec.~\ref{s4} we make some brief concluding remarks.

\section{An exactly solvable nonanalytic potential}
\label{s2}
In this section we consider the $\cP\cT$-symmetric Hamiltonian
\begin{equation}
H=p^2+ix|x|,
\label{e6}
\end{equation}
whose potential is a nonanalytic function of $x$. The Schr\"odinger eigenvalue
differential equation associated with this Hamiltonian is
\begin{equation}
\left(-\frac{d^2}{dx^2}+ix|x|-E\right)\psi(x)=0,
\label{e7}
\end{equation}
where $x$ is real and where the eigenfunction $\psi(x)$ is required to obey the
boundary conditions that $\psi\to0$ as $x\to\pm\infty$.

To solve this differential equation, we partition the real axis into two
regions. In the region $x>0$ the differential equation (\ref{e7}) takes the form
\begin{equation}
\left(-\frac{d^2}{d x^2}+ix^2-E\right)\psi(x)=0
\label{e8}
\end{equation}
and the exact solution is
\begin{eqnarray}
\psi(t)=c_1D_\nu\left(xe^{i\pi/8}\sqrt{2}\right)+c_2D_\nu\left(-xe^{i\pi/8}
\sqrt{2}\right).
\label{e9}
\end{eqnarray}
Here, $D_\nu$ is the parabolic cylinder function with
\begin{equation}
\nu=\half Ee^{-i\pi/4}-\half,
\label{e10}
\end{equation}
and $c_1$ and $c_2$ are arbitrary constants. The boundary condition $\lim_{
x\to+\infty}\psi(x)=0$ implies that $c_2=0$. Thus, for $x>0$ we have
\begin{equation}
\psi(x)=c_1D_\nu\left(xe^{i\pi/8}\sqrt{2}\right).
\label{e11}
\end{equation}

Similarly, in the region $x<0$ the differential equation (\ref{e7}) becomes
\begin{equation}
\left(-\frac{d^2}{d x^2}-ix^2-E\right)\psi(x)=0,
\label{e12}
\end{equation}
whose exact solution is
\begin{equation}
\psi(s)=d_1D_\mu\left(xe^{-i\pi/8}\sqrt{2}\right)+d_2D_\mu\left(-xe^{-i\pi/8}
\sqrt{2}\right).
\label{e13}
\end{equation}
Here,
\begin{equation}
\mu=\half e^{i\pi/4}E-\half
\label{e14}
\end{equation}
and $d_1$ and $d_2$ are arbitrary constants. The boundary condition $\lim_{
x\to-\infty}\psi(x)=0$ implies that $d_1=0$. Thus, for $x<0$ we have
\begin{equation}
\psi(x)=d_2D_\mu\left(-xe^{-i\pi/8}\sqrt{2}\right).
\label{e15}
\end{equation}

We must patch the two solutions (\ref{e11}) and (\ref{e15}) together at the
origin $x=0$. Continuity of $\psi(x)$ at $x=0$ implies that
\begin{equation}
c_1D_\nu(0)=d_2D_\mu(0),
\label{e16}
\end{equation}
and continuity of $\psi'(x)$ at $x=0$ implies that
\begin{equation}
\label{e17}
c_1e^{i\pi/8}D'_\nu(0)=-d_2e^{-i\pi/8}D'_\mu(0).
\end{equation}
Taking the ratio of (\ref{e17}) and (\ref{e16}) eliminates the constants $c_1$
and $d_2$ and gives the following exact equation for the eigenvalues:
\begin{equation}
\frac{e^{i\pi/8}D'_\nu(0)}{D_\nu(0)}=-\frac{e^{-i\pi/8}D_\mu'(0)}{D_\mu(0)}.
\label{e18}
\end{equation}
This condition can be rewritten simply in terms of Gamma functions as
\begin{equation}
e^{i\pi/8}\frac{\Gamma\left(\tquat-\quat Ee^{-i\pi/4}\right)}{\Gamma\left(\quat
-\quat Ee^{-i\pi/4}\right)}+e^{-i\pi/8}\frac{\Gamma\left(\tquat-\quat Ee^{i\pi/
4}\right)}{\Gamma\left(\quat-\quat Ee^{i\pi/4}\right)}=0.
\label{e19}
\end{equation}
As required by $\cP\cT$ symmetry, the secular equation (\ref{e19}) is a {\it
real} function of $E$. This is so because it is the sum of two terms that are
complex conjugates of each other.

\begin{figure*}[t!]
\vspace{3.3in}
\includegraphics{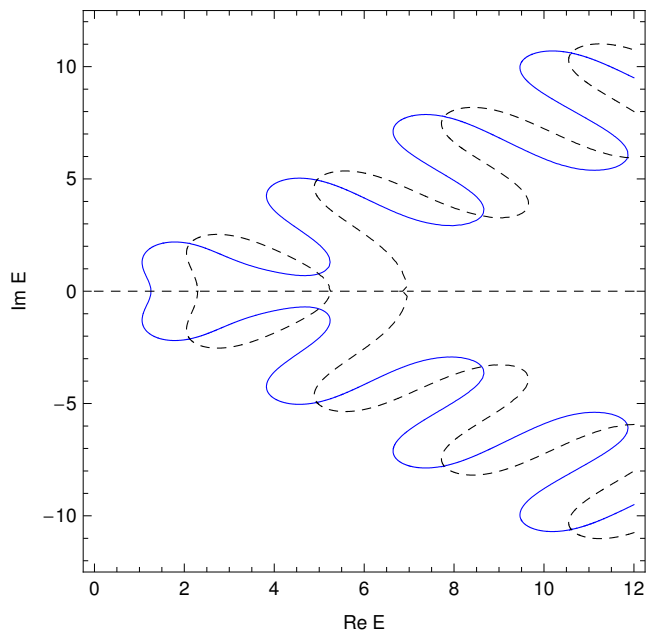}
\caption{Numerical solution to the secular equation (\ref{e19}) for the
eigenvalues of the Hamiltonian $H=p^2+ix|x|$. The solid line is the region in
the complex-$E$ plane where the real part of the secular determinant vanishes.
The dotted line indicates where the imaginary part of the secular equation
vanishes. The intersections of the solid and dotted lines are the eigenvalues.
There is one real eigenvalue, which is located at $E=1.258092$. All other
eigenvalues come in complex-conjugate pairs.}
\label{f1}
\end{figure*}

To solve (\ref{e19}) for $E$, we substitute $E={\rm Re}\,E+i\,{\rm Im}\,E$ and
take the real and imaginary parts of the resulting equation. We then plot in
Fig.~\ref{f1} the curves in the complex-$E$ plane along which the real part of
(\ref{e19}) vanishes (solid line) and the imaginary part of (\ref{e19}) vanishes
(dotted line). [Of course, the condition of $\cP\cT$ symmetry requires that the
dotted line lie along the real-$E$ axis. However, note that the real-$E$ axis is
not the only curve along which the imaginary part of (\ref{e19}) vanishes.]

The intersections of the solid and dotted lines in Fig.~\ref{f1} are the
eigenvalues of $H$ in (\ref{e6}). Note that there is only one real eigenvalue;
all other intersections occur in complex-conjugate pairs. The numerical value of
the real eigenvalue is
\begin{equation}
E_0=1.258092\ldots\,.
\label{e20}
\end{equation}

\section{Numerical study of a class of nonanalytic potentials}
\label{s3}
In this section we examine numerically the eigenvalue differential equation for
the complex $\cP\cT$-symmetric non-Hermitian Hamiltonian
\begin{equation}
H=p^2+(ix)^a|x|^b,
\label{e21}
\end{equation}
where $a$ and $b$ are real parameters.

We begin by determining the appropriate $\cP\cT$-symmetric boundary conditions
to be imposed on the eigenfunctions of $H$ in (\ref{e21}). Using WKB analysis,
we find the possible asymptotic behaviors of the eigenfunction $\psi(x)$
satisfying the time-independent Schr\"odinger equation
\begin{equation}
\label{e22}
\left(-\frac{d^2}{d x^2}+(ix)^a|x|^b-E\right)\psi(x)=0.
\end{equation}
For example, when $x>0$, the controlling factor of the asymptotic behavior of
$\psi(x)$ as $x\to+\infty$ is
\begin{equation}
\exp\left[\pm\frac{2}{a+b+2}i^{a/2}x^{(a+b+2)/2}\right].
\label{e23}
\end{equation}
Thus, for $a<2$ there exists a solution that grows exponentially and another
that decays exponentially for large positive $x$. The same is true for large
negative $x$ so long as $a<2$. To determine the eigenvalues for $a<2$ we impose
the boundary conditions that $\psi(x)\to0$ as $|x|\to\infty$ on the real-$x$
axis. Note that because the potential is not an analytic function of $x$, the
notion of Stokes' wedges in the complex-$x$ plane in which the boundary
conditions are imposed is not applicable.

\begin{figure*}[ht!]
\vspace{6.5in}
\includegraphics{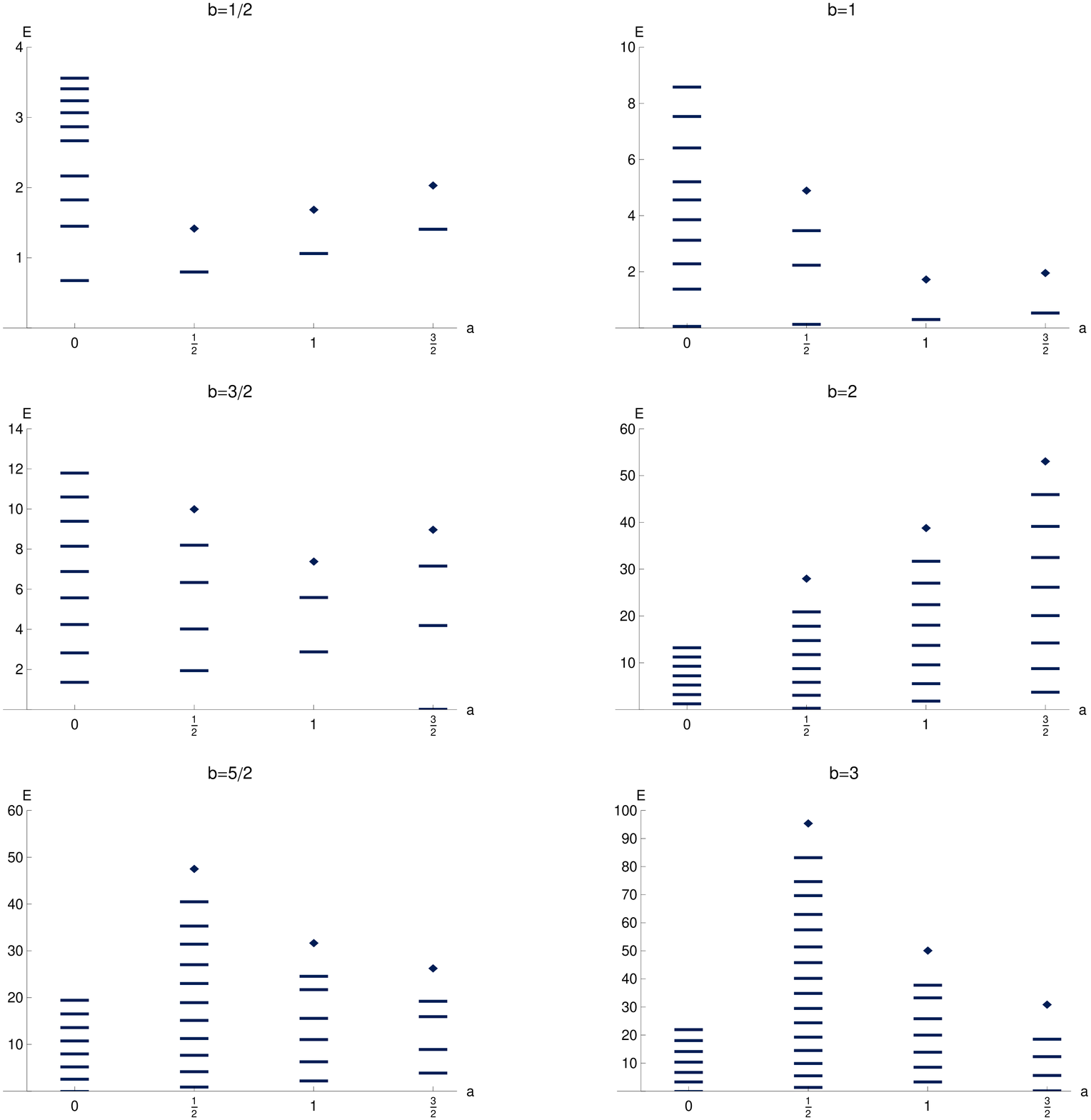}
\caption{Real eigenvalues for the Hamiltonian $H=p^2+(ix)^a|x|^b$ in
(\ref{e21}). The data for this figure is taken from Tables \ref{t1} and
\ref{t2}. The symbol $\blacklozenge$ indicates that there are no more real
eigenvalues in the tower. Notice that the number of real eigenvalues increases
with increasing $b$ and decreases with increasing $a$.}
\label{f2}
\end{figure*}

We have calculated the eigenvalues for various values of $a$ and $b$ and our
results are listed in Tables \ref{t1} and \ref{t2} and plotted in Fig.~\ref{f2}.
Clearly, when $b$ is an even integer, the Hamiltonian in (\ref{e21}) reduces to
that in (\ref{e1}) and it has an entirely real spectrum. However, for other values
of $b$, when $a\neq0$ there are only a finite number of real eigenvalues. The
number of real eigenvalues decreases as $a$ increases, and increases as $b$
increases.

\begin{table}[!h]
\begin{center}
\begin{tabular}{|c||c|c|c|c|c|c|}\hline
$E_i$ & $b=1/2$ & $b=1$ & $b=3/2$ & $b=2$ & $b=5/2$ & $b=3$\\\hline
$a=0$ & 1.059617 & 1.01879 & 1.00118 & 1 & 1.00859 & 1.02295\\
$$    & 1.833394 & 2.33811 & 2.70809 & 3 & 3.24223 & 3.45056\\
$$    & 2.210015 & 3.24820 & 4.17714 & 5 & 5.72682 & 6.37029\\
$$    & 2.550647 & 4.08795 & 5.58566 & 7 & 8.31328 & 9.52208\\
$$    & 3.051182 & 4.82010 & 6.92282 & 9 & 10.9916 & 12.8703\\
$$    & 3.253157 & 5.52056 & 8.22687 & 11 & 13.7342 & 16.3694\\
$$    & 3.452132 & 6.16331 & 9.49059 & 13 & 16.5353 & 20.0009\\
$$    & 3.623138 & 7.37218 & 10.7317 & 15 & 19.3837 & 23.7455\\
$$    & 3.793400 & 8.48849 & 11.9453 & 17 & 22.2757 & 27.5924\\
$$    & 3.943821 & 9.53546 & 13.1419 &  19 & 25.2052 & 31.5308\\
$$    & \ldots   & \ldots  & \ldots  & \ldots & \ldots & \ldots\\\hline
$a=1/2$ & 1.180777 & 1.08693 & 1.05583 & 1.04896 & 1.05404 & 1.06568\\
$$ & $\blacklozenge$ &3.19578 & 3.27843 & 3.43454 & 3.59460 & 3.74791\\
$$ & & 4.4220 & 5.36421 & 6.05174 & 6.64515 & 7.17496\\
$$ & & $\blacklozenge$ & 7.67568 & 8.79101 & 9.91884 & 10.9735\\
$$ & & & 9.53919 & 11.6207 & 13.4256 & 15.1112\\
$$ & & & $\blacklozenge$ & 14.5219 & 17.0514 & 19.4889\\
$$ & & & & 17.4829 & 20.8691 & 24.1139\\
$$ & & & & 20.4952 & 24.7239 & 28.9111\\
$$ & & & & 23.5529 & 28.8137 & 33.9218\\
$$ & & & & 26.6504 & 32.7868 & 39.0482\\
$$ & & & & \ldots & 37.2141 & 44.3936\\
$$ & & & &   & 41.0803 & 49.7770\\
$$ & & & &   & 46.2256 & 55.4476\\
$$ & & & &   & $\blacklozenge$ & 60.9963\\
$$ & & & &   &  & 67.0561\\
$$ & & & &   &  & 72.5848\\
$$ & & & &   &  & 79.2958\\
$$ & & & &   &  & 84.3126\\
$$ & & & &   &  & 92.7345\\
$$ & & & &   &  & $\blacklozenge$\\\hline
\end{tabular}
\end{center}
\caption{\label{t1} Real eigenvalues for the Hamiltonian $H=p^2+(ix)^a|x|^b$ in
(\ref{e21}), where $a=0$ and $a=1/2$. The symbol $\blacklozenge$ means that
there are no more real eigenvalues, while $\ldots$ indicates that the spectrum
is entirely real.}
\end{table}

\begin{table}[!h]
\begin{center}
\begin{tabular}{|c||c|c|c|c|c|c|}\hline
$E_i$ & $b=1/2$ & $b=1$ & $b=3/2$ & $b=2$ & $b=5/2$ & $b=3$\\\hline
$a=1$ & 1.446448 & 1.25809 & 1.18627 & 1.15627 & 1.14615 & 1.14685\\
$$ & $\blacklozenge$ & $\blacklozenge$ & 4.21683 & 4.10923 & 4.13051 & 4.19436\\
$$ & &  & 6.93323 & 7.56227 & 7.95153 & 8.30206\\
$$ & &  & $\blacklozenge$ & 11.3144 & 12.0844 & 12.9101\\
$$ & &  & & 15.2916 & 16.8072 & 18.1062\\
$$ & &  & & 19.4515 & 21.3065 & 23.5322\\
$$ & &  & & 23.76667 & 27.4779 & 29.6147\\
$$ & &  & & 28.2175 & 30.3268 & 35.3873\\
$$ & &  & & 32.7891 & $\blacklozenge$ & 42.9034\\
$$ & &  & & 37.4698 &  & 47.4048\\
$$ & &  & & \ldots &  & $\blacklozenge$\\\hline
$a=3/2$ & 1.791941 & 1.48873 & 1.36338 & 1.30151 & 1.26993 & 1.2550\\
$$ & $\blacklozenge$ & $\blacklozenge$ & 5.52801 & 4.96979 & 4.80096 & 4.7494\\
$$ & &  & 8.50818 & 9.48003 & 9.60759 & 9.7042\\
$$ & &  & $\blacklozenge$ & 14.5305 & 14.6672 & 15.2406\\
$$ & &  &  & 19.9977 & 21.7069 & 21.891\\
$$ & &  &  & 25.8103 & 24.9567 & 28.1147\\
$$ & &  &  & 31.9205 & $\blacklozenge$ & $\blacklozenge$ \\
$$ & &  &  & 38.2938 &  & \\
$$ & &  &  & 44.904 &  & \\
$$ & &  &  & 51.7304 &  & \\
$$ & &  &  & \ldots &  & \\
\hline
\end{tabular}
\end{center}
\caption{\label{t2} Same as in Table \ref{t1} except that $a=1$ and $a=3/2$.}
\end{table}

\section{Concluding remarks}
\label{s4}

Most $\cP\cT$-symmetric potentials $V(x)$ studied so far in the literature are
special because they are analytic. In this paper we have explored a new class of
nonanalytic $\cP\cT$-symmetric potentials of the form $V(x)=(ix)^a|x|^b$, and on
the basis of numerical and theoretical calculations we are led to conjecture that,
except in rare cases, analyticity is an essential feature that is necessary for
the Hamiltonian to have a real spectrum.

\vspace{0.5cm}
\footnotesize
\noindent
We thank D.~C.~Brody for helpful discussions. CMB is supported by a grant from
the U.S.~Department of Energy. DWH receives financial support from Symplectic Ltd.

\end{document}